\documentclass[preprint,showpacs,preprintnumbers]{revtex4}
\usepackage{amsmath}
\usepackage{graphicx}
\usepackage{dcolumn}
\usepackage{braket}
\usepackage{bm}

\begin{document}

\title{Polaron picture of the two-photon quantum Rabi model}
\author{Lei Cong$^{1}$, Xi-Mei Sun$^{1}$, Maoxin Liu$^{2}$, Zu-Jian Ying$^{2,3,*}$, and Hong-Gang Luo$^{1,2,*}$}
\address{
$^{1}$ Center for Interdisciplinary Studies and Key Laboratory for Magnetism and Magnetic Materials of the MoE, Lanzhou University, Lanzhou 730000, China \\
$^{2}$ Beijing Computational Science Research Center, Beijing 100084, China\\
$^{3}$ CNR-SPIN and Dipartimento di Fisica ``E. R. Caianiello'', Universit\`a di Salerno, I-84084 Fisciano (Salerno), Italy
 }
\date{\today }

\begin{abstract}
We employ a polaron picture to investigate the properties of the two-photon quantum Rabi model (QRM), which describes a two-level or spin-half system coupled with a single bosonic mode by a two-photon process. In the polaron picture, the coupling in the two-photon process leads to spin-related asymmetry so that the original single bosonic mode splits into two separated frequency modes for the opposite spins, which correspond to two \textit{bare} polarons. Furthermore, the tunneling causes these two bare polarons to exchange their components with each other, thus leading to additional \textit{induced} polarons. According to this picture, the variational ground-state wave function of the two-photon QRM can be correctly constructed, with the ground-state energy and other physical observables in good agreement with the exact numerics in all the coupling regimes. Furthermore, generalization to multiple induced polarons involving higher orders in the tunneling effect provides a systematic way to yield a rapid convergence in accuracy even around the difficult spectral collapse point. In addition, the polaron picture provides a distinctive understanding of the spectral collapse behavior, that is about the existence of discrete energy levels apart from the collapsed spectrum  at the spectral collapse point. This work illustrates that the polaron picture is helpful to capture the key physics in this nonlinear light-matter interaction model and indicates that this method can be applicable to more complicated QRM-related models.
\end{abstract}

\maketitle

\section{Introduction}\label{intro}
Two-photon coupling in light-matter interaction is a nonlinear process  in  which  electron  transition between quantum levels occurs through the simultaneous emission or absorption of two photons \cite{ANDP:ANDP19314010303,Sorokin:1964:TAP:1662395.1662404,Prokhorov828}. It is usually weaker than the related first order processes but may be enhanced by advanced quantum technologies\cite{PhysRevA.92.033817,PhysRevA.95.063844,PhysRevA.97.023624,PhysRevA.97.013851}. Its wide relevance can be found in quantum optics \cite{Edamatsu2004,Barz2010,PhysRevLett.23.880}, quantum communication \cite{PhysRevA.67.062309}, atomic physics \cite{PhysRev.113.179}, semiconductor physics \cite{Hayat2008}, astrophysics \cite{PhysRevA.24.183}, biophysics \cite{Zong2017,Denk73}, and so on.

One of the simplest models describing this type of interaction is called two-photon quantum Rabi model (QRM), which describes a two-level or spin-half system nonlinearly coupled to a single cavity mode. It is a direct extension of the standard QRM \cite{PhysRev.49.324,PhysRev.51.652} which involves one-photon absorption and emission. The full nonlinear QRM has a linear and a nonlinear coupling term \cite{yingcong}, which is the appropriate description in some situations, see e.g. \cite{PhysRevLett.120.160403}. Here we only focus on the physics of the two-photon process, thus we consider the two-photon QRM without the linear coupling term.

The two-photon QRM has been realized in many different experimental systems, for examples, Rydberg atoms \cite{PhysRevLett.59.1899,PhysRevLett.88.143601} and quantum dots \cite{PhysRevB.73.125304,PhysRevB.81.035302,PhysRevB.94.085309,PhysRevLett.120.213901} in microwave cavities.
Usually, the coupling strength is weak in the experimental setups mentioned above. However, it is believed that all relevant parameter regimes including the strong coupling could be
reached by using trapped-ion technologies \cite{PhysRevA.92.033817,PhysRevA.95.063844,PhysRevA.97.023624}. Very recently, it was also proposed that the two-photon QRM could be naturally implemented in an undriven system using superconducting circuit \cite{PhysRevA.97.013851, simonenew}, covering a wide coupling regime, which may inspire further interest on this model. Besides the experimental setups, already realized or proposed, the two-photon QRM is also of considerable interest for realistic applications including two-photon lasers \cite{PhysRevLett.47.171,PhysRevB.81.035302} and two-photon optical bistability \cite{PhysRevA.24.2029}.

The theoretical study of the two-photon QRM shows a particular feature \cite{0295-5075-42-1-001,PhysRevA.92.033817,1751-8121-49-46-464002,PhysRevA.95.053854}, often referred to as the spectral collapse, which appears when the qubit-cavity coupling strength $g$ becomes comparable to the bosonic mode frequency $\omega$. Exact numerical results indicate that the model does not have eigenstates in the Hilbert space spanned by the photon number states when the coupling strength is larger than the frequency, i.e., $|2g|>\omega$ \cite{0295-5075-42-1-001} (here we set $\hbar =1$). Even though some attention has been paid to the problem of the spectral collapse, the characteristic behavior of eigen-functions at the spectral collapse point $|2g| = \omega$ has not been completely understood \cite{PhysRevA.92.033817}.

In the regime of $|2g| \le \omega$, the analytical solution of the two-photon QRM has been obtained in several stages. During the early stage, similar to the Juddian points in QRM \cite{0022-3719-12-9-010}, only some special points were known to be exactly solvable \cite{0305-4470-35-39-307}. In 2012, a complete analytical solution of the two-photon QRM \cite{PhysRevA.85.043805} has been obtained by using Braak's approach \cite{PhysRevLett.107.100401}, followed by \cite{1751-8121-49-46-464002,PhysRevA.86.023822,Peng2013,PhysRevA.91.037802,1751-8121-50-24-244003}. However, for these solutions, only the energy spectrum is exactly available, it is still difficult to calculate the explicit wave function which is necessary to calculate many other physical observables apart from the energy.

Some approximate methods have been proposed to explore the properties of the model \cite{1751-8121-49-46-464002,PhysRevA.93.043814,PhysRevA.37.2683}. However, the approximate methods employed often do not produce adequate results in the entire range of parameters, especially the regime around $g/\omega \sim 0.5$ proves difficult to access. For example, in the finite-dimensional approximation method \cite{1751-8121-49-46-464002}, an abnormal upward trend of the ground-state energy has been observed. The perturbative approach was found to be working well only in the weak coupling regime \cite{PhysRevA.93.043814}. The rotating wave approximation has also been applied to the two-photon QRM \cite{PhysRevA.37.2683,PhysRevA.49.473,PhysRevA.84.042110}. However, the reduced model after the RWA shows qualitative differences from the full two-photon QRM \cite{0295-5075-42-1-001}, as the interesting spectral collapse is missing in the reduced model.

Thus, any method that provides a clear physical picture for the whole coupling regime and maintains a high accuracy deserves an effort to be developed for and applied to the two-photon QRM.

Recently, a variational polaron picture has been successfully applied to the standard QRM \cite{PhysRevA.92.053823,PhysRevA.95.063803} as well as the spin-boson model  \cite{PhysRevB.89.121108, PhysRevB.90.075110}. This method shows its high accuracy and high efficiency in exploring the whole coupling regime of the model, including the phase transition or crossover regime \cite{PhysRevA.92.053823, PhysRevA.95.063803} which is difficult for many other existing approximations, such as the widely used generalized rotating-wave approximation (GRWA) \cite{PhysRevLett.99.173601, PhysRevA.83.065802, PhysRevA.94.063824, PhysRevA.86.015803}. Moreover, the polaron picture can provide insights into the underlying physics because the effect of the tunneling term is specifically considered \cite{PhysRevA.92.053823,PhysRevB.89.121108}.

In this work, the study of the polaron picture for the two-photon QRM provides a clear intuition for several physical aspects of the model. First of all, the polaron picture shows that the two-photon interaction leads to a spin-related asymmetry so that the original single bosonic mode splits into two separated frequency modes. This is an intriguing difference from the case of the standard QRM. Tunneling will cause further deformation of these two asymmetric modes. Secondly, for the ground state, in addition to the polarons connected with the asymmetry modes, there also exist some tunneling induced polarons. These induced polarons are important to guarantee the accuracy of the variational state when the coupling increases from the weak to strong regime, especially close to the spectral collapse. By comparison with exact diagonalization (ED), we show that our method based on the polaron picture yields a high accuracy not only for the ground-state energy but also for the wave function profile and various physical observables. Thirdly, it is found from the ED result that the spectral collapse is actually not complete, i.e. discrete energy levels coexist with collapsed energy levels. The polaron picture shows that the reason these bound states exist is that the tunnelling can induce an effective potential well with finite depth. The number of bound states depends on the strength of the tunneling $\Omega$. It is expected that the discrete bound states disappear as $\Omega \rightarrow 0$. These results demonstrate that the polaron picture is able to capture the essential physics involved in the light-matter interaction models not only for linear coupling \cite{PhysRevA.92.053823} but also for the nonlinear processes we study here.

The paper is organized as follows. In Sec. \ref{model}, the two-photon QRM model Hamiltonian is introduced. In Sec. \ref{polaron-picture}, in order to introduce the polaron picture, the model is first reformulated in terms of position and momentum operators so that the  single-particle potential is obtained. Then, we show how the tunneling modifies the total effective potential. Finally, the variational ground state is constructed based on the polaron picture. In Sec. \ref{results}, we further examine the profile of the wave function and calculate several other physical observables. Those results show the accuracy of the polaron state. And, for a systematical consideration of the methodology itself, we also extend our analysis to a multi-polaron wave function Ansatz. In Sec. \ref{Discussion}, we compare the polaron picture for the one-photon QRM and the two-photon QRM, and their similarities and differences are discussed. In particular, we provide a novel perspective to understand the behavior of the spectrum at the collapse point, as facilitated by the polaron picture. Finally, Sec. \ref{conclusion} is devoted to a brief summary.

\section{The two-photon QRM model}\label{model}

The Hamiltonian of the two-photon QRM reads ($\hbar = 1$)
\begin{equation}\label{Hamiltonian}
H_c=\omega a^\dag a +\frac{\Omega}{2}\sigma_z +g\sigma_x{\left[(a^\dag)^2 + a^2\right]},
\end{equation}
where $\Omega $ is the energy level splitting of the two level system or spin-half qubit, $\sigma_{x,z}$ are the Pauli matrices to describe the qubit, $\hat{a}^{\dagger}$ and $\hat{a}$ are the bosonic creation and annihilation operators, respectively, of the bosonic mode with frequency $\omega$, and $g$ denotes the coupling strength between the two level system and the bosonic mode.

It is convenient to rewrite the two-photon QRM in the spin-boson representation \cite{1751-8121-49-46-464002}, obtained through a unitary transformation of the Hamiltonian by the operator  $\text{e}^{-\frac{i\pi}{4} \sigma_{y}}$. The Hamiltonian then takes the following form:
\begin{equation}\label{Hamiltonian1}
H=\omega a^\dag a +\frac{\Omega}{2}\sigma_x +g\sigma_z{\left[(a^\dag)^2 + a^2\right]}.
\end{equation}
In this form the $\Omega$ term effectively plays a role of tunneling rate
between the spin-up and spin-down states. The terminology of tunneling we adopt here is also used in \cite{PhysRevB.89.085421,PhysRevA.92.053823,Feng2013,PhysRevB.89.085421}. And the spin notation is defined according to the Hamiltonian \eqref{Hamiltonian1} to avoid ambiguity in the following discussion. Besides, for general discussion, the value of $g$ can be negative or positive. Without loss of generality, in further discussion we assume positive $g$, as $\sigma_z$-independent physical quantities including the energy, the mean photon number and the tunneling strength are even functions of $g$, while the $\sigma_z$-linearly-dependent quantities like the coupling correlation is an odd function of $g$ and the analysis of the effective potential for negative $g$ is the same except for the sign exchange in the spin $\sigma_z$.

\section{Polaron picture analysis}\label{polaron-picture}
In this section, we analyse the polaron picture of the two-photon model in order to introduce a clear physical understanding. Such a polaron picture contributes to our understanding of this model as well as to the construction of ground-state wave function of the system.

In terms of the quantum harmonic oscillator with dimensionless formalism $\hat{a}^\dag=(\hat{x}-i\hat{p})/\sqrt{2}$, $\hat{a}=(\hat{x}+i\hat{p})/\sqrt{2}$, where $\hat{x}=x$ and
$\hat{p}=-i \frac{\partial}{\partial x}$ are the position and momentum operators respectively, the model can be rewritten as
\begin{equation} \label{Hamiltonian2}
H =\sum_{S_z=\pm} {\left(h^{S_z}{\ket{S_z}\bra{S_z}}+\frac{\Omega}{2} {\ket{S_z}\bra{\bar{S_z}}}\right)}+\varepsilon_0
\end{equation}
where $+(-)$ labels
the spin up (down) state in the z-direction and $\bar{S_z}=-{S_z}$.
$\varepsilon_0=-\omega/2$ is a constant. The single-particle Hamiltonian apart from the tunneling is denoted by $h^{\pm}=\frac{\omega}{2}(1\mp2g')(\hat{p}^2+v^{\pm})$, where $v^{\pm}$ is defined by
\begin{equation}\label{vpm}
v^{+}=\frac{1+2g'}{1-2g'}\hat{x}^2, \hspace{0.3cm} v^{-}=\frac{1-2g'}{1+2g'}\hat{x}^2,
\end{equation}
and here $g'=g/\omega$. $v^{+} (v^{-})$ defines a \textit{bare} potential. One notes that the bare potential $v^{+}$ would diverge if $g' \rightarrow 1/2$. However, this divergence would not lead to difficulty for the whole model because there is a factor $(1-g')$ in the front of $v^+$ inside of the $h^+$ which cancels the divergence. Based on the above mentioned transformation, we will further discuss the effects of coupling and tunneling of the system step-by-step in the following.


\begin{figure}[tbp]
  \includegraphics[width=0.6\columnwidth]{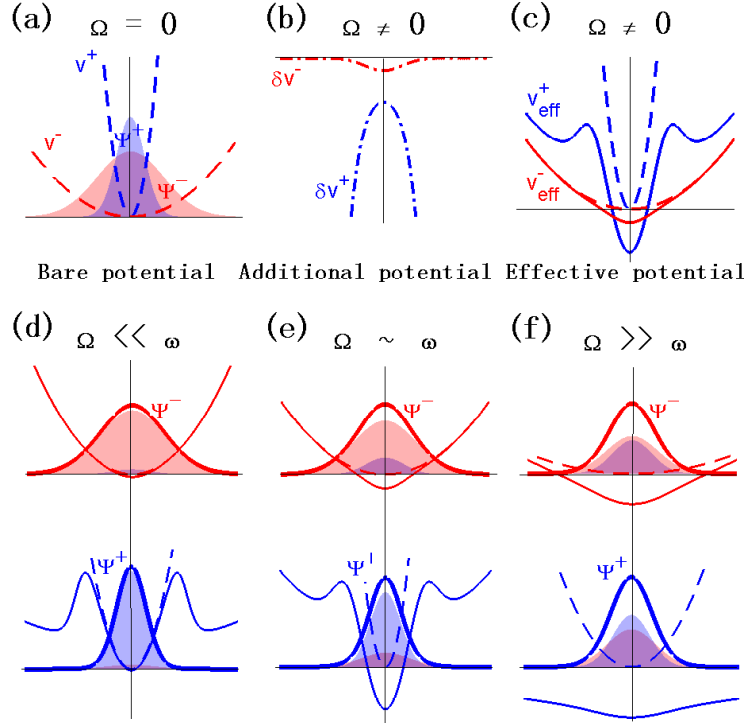}
  \caption{(Color online)  Top graph (a)-(c): Schematic diagram for bare potentials and tunneling-induced deformation of effective potentials. (a) In the absence of the tunneling between two spin states, i.e. $\Omega=0$, the single bosonic mode is nonlinearly coupled with two spin states labeled by $\pm$ to form two \textit{bare} potentials $v^{\pm}$ which determine two \textit{bare} polarons ($\Psi^{-}$ and $\Psi^{+}$ filled in blue and red). (b) When the tunneling is switched on, namely, $\Omega \neq 0$, the \textit{bare} polarons couple with each other and induce an additional potential $\delta v^+=-\frac{\Omega}{(1-2g')\omega}\frac{\Psi^{-}}{\Psi^{+}}$ and $\delta v^-=-\frac{\Omega}{(1+2g')\omega}\frac{\Psi^{+}}{\Psi^{-}}$. (c) The total effective potential $v^\pm_\mathrm{eff}=v^{\pm}+\delta v^{\pm}$. The solid lines denote the effective potentials and the dashed lines are the \textit{bare} potentials.
      Bottom graph (d)-(f):   Tunneling-induced polarons in different regimes of tunneling strength. In the presence of tunneling, each wave function $\Psi^{\pm}$ (sketched by thick solid lines) consists of two separated components, one originating from the bare polaron and the other induced by the tunneling, depicted respectively by $\Psi^{+}(x)=\alpha_1\varphi_{\alpha_1}+\alpha_2\varphi_{\alpha_2}$ and  $\Psi^{-}(x)=\beta_1\varphi_{\beta_1}+\beta_2\varphi_{\beta_2}$ [see Eq. (\eqref{twocpolaron})]. The wave function $\Psi^{-}(x)$ consists of an original polaron with low frequency (filled in red) and an \textit{induced} polarons with high frequency (filled in blue). Meanwhile, the wave function $\Psi^{+}(x)$ consists of an original polaron with high frequency (filled in blue) and an \textit{induced} polaron with low frequency (filled in red).
Panels (d-f) are representative cases in the weak, intermediate, and strong tunneling regimes respectively, which shows successively the increasing weight for the \textit{induced} polarons.
} \label{fig1}
\end{figure}

\subsection{Effective potential picture}\label{potential}
\subsubsection{Bare potential}
We first consider the case without tunneling, namely, $\Omega=0$. In this case, the photonic term $\omega a^\dag a $ and coupling term $g\sigma_z{\left[(a^\dag)^2 + a^2\right]}$ are combined together to define the bare potential $v^{-} (v^{+})$ [see Eq. (\ref{vpm})], which is a harmonic oscillator well in a lower (higher) frequency mode for the spin-down (spin-up) component. $v^{-} (v^{+})$ are sketched by the red (blue) dashed lines in Fig. \ref{fig1}(a). Each of them accommodates a \textit{bare} polaron \cite{PhysRevA.92.053823,PhysRevA.95.063803,PhysRevB.89.121108, PhysRevB.90.075110} as sketched by the wave packet $\Psi^{\pm}$ in the shaded area in Fig. \ref{fig1}(a).

For $g'= 0$, the bare potentials ($v^{\pm}$) of the two spin components are the same. However, for $g'\neq 0$ but $g' < 1/2$, $v^{\pm}$ separates from each other, as recognized from Eq. (\eqref{vpm}). Therefore the bare polarons associated with the opposite spins are also not symmetric any more.  At this point, one immediately sees an interesting result. The interaction via two-photon processes leads to a spin asymmetry so that opposite spins have different effective mass (defined by the kinetic energy in the form of $\hat{p}^2/(2m^*)$, with $m^*$ being the effective mass) and different potentials, with a tendency of the frequency increasing and decreasing for spin-up and spin-down components, respectively. This is in contrast to the spin symmetry in the standard QRM \cite{PhysRevA.92.053823,PhysRevA.95.063803} with single-photon coupling where the two spins have the same effective mass and the effective potential is symmetric with respect to the spatial origin under spin exchange.

 For $g'=1/2$, $v^{-}$ vanishes. Thus, this lower frequency mode behaves like a free particle. This corresponds to the spectral collapse, as also discussed in the previous study \cite{PhysRevA.92.033817}. Such a simple picture explains the reason why the collapse point ($g=\omega/2$) is only related to the field frequency $\omega$, and can also be extended to understand the same collapse point in  the two-photon Dicke model \cite{PhysRevA.95.053854}. More discussion about the spectral collapse behavior can be seen in Sec. \ref{SpectralCollapse}.

\subsubsection{Tunneling-induced additional potential}

If $\Omega\neq0$,  the channel of the tunneling from the term $\frac{\Omega}{2}\sum_{S_z=\pm} {\ket{S_z}\bra{\bar{S_z}}}$ in Eq. (\ref{Hamiltonian2}) is turned on. The tunneling will couple the two bare potentials and deform the potential profiles, thus effectively leading to additional potentials $\delta v^{\pm}$, as shown in Fig. \ref{fig1}(b). Indeed, we can obtain these additional potentials from the Sch\"odinger equation $H|G\rangle = E|G\rangle$, where $|G\rangle$ [see the form in Eq. (\ref{trial})] is the assumed eigen-state wave function and $E$ is the eigen-energy. The Sch\"odinger equation can be rewritten in the single-particle form
\begin{equation}
\frac{1}{2}\omega(1\mp2g')(\hat{p}^2+v^{\pm}+\delta v^{\pm}){\Psi^{\pm}}=E{\Psi^{\pm}}, \label{sch2}
\end{equation}
where $\delta v^{\pm}= -\frac{\Omega}{(1{\mp}2g')\omega}\frac{\Psi_{\mp}}{\Psi^{\pm}}$ is  the additional potential (dash-dotted lines in Fig. \ref{fig1}(b)) induced by the tunneling between the bare potentials. As shown in Fig. \ref{fig1}(c), combination of the additional potential with the original bare potential gives rise to the effective potential:
\begin{equation}\label{barepotential}
v^\pm_\mathrm{eff} = v^\pm + \delta v^\pm.
\end{equation}
As a result, an effective potential deformed from the bare potential is present. Similar discussions about the tunneling term in QRM can be found in Refs. \cite{PhysRevA.92.053823, PhysRevB.89.085421}.

Previously, we saw that $v^-$ becomes flat and $v^+$ diverges when $g' \rightarrow 1/2$. Therefore, one can expect that the corresponding wave function $\Psi^{-}$ ($\Psi^{+}$) will collapse (diverge). Here, when the tunneling is switched on, both $v^-$ and $v^+$ are deformed by an additional $\delta v^{\pm}$, see Eq. (\ref{barepotential}). As a result, the corresponding wave function $\Psi^{-}$ ($\Psi^{+}$) may no longer collapse or diverge. Exact numerical result obtained by the exact diagonalization (ED) method confirms the above conjecture, see Appendix \ref{appa}.

The above effective potential picture provides a simple intuition, which can be addressed below: (i) The nonlinear-interaction-driven spin asymmetry in the frequency space is a new feature in this model while in the standard QRM this kind of spin asymmetry is not present. And the interesting spectral collapse is due to a flat potential. (ii) The tunneling process in deforming the effective potential as well as inducing new changes of the eigenstate are key for the physics of the two-photon QRM. In the following we will analyse the variational ground state based on the above polaron picture.

\subsection{Ground-state wave function Ansatz}\label{method}

The wave function of the two-photon QRM in the position representation can be written generally as
\begin{equation}\label{trial}
\ket G=\frac{1}{\sqrt{2}}\big(\Psi^{+}(x)\ket\uparrow-\Psi^{-}(x)\ket\downarrow \big)
\end{equation}
where $\Psi^{\pm}$ is the wave function associated with the spin state $|S_z\rangle$. The ground state in the variational method is determined by the minimization of the energy with respect to the trial variational states. Here we assume that $\Psi^{\pm}$ are real functions.

\subsubsection{Variational state with polaron and tunneling-induced polarons (N=2)}

 According to Eq. (\ref{Hamiltonian2}), the effect of the tunneling term is to couple the higher and lower frequency modes. One can imagine that a direct consequence of the tunneling process is that the two bare polarons will exchange part of their components. In this situation, the higher frequency mode will contribute a component in the lower frequency mode, and meanwhile, the lower frequency mode also leave a component in the higher frequency one, see Fig. \ref{fig1}(e-f). These additional components can be considered as polarons induced by the tunneling. In order to capture this picture, it is very natural to consider the ground-state wave function consisting of two pairs of polaron, as given by
\begin{equation}\label{twocpolaron}
\begin{aligned}
\left\{ \begin{array} {ll}
\Psi^{+}(x)&=\alpha_1\varphi_{\alpha_1}(\xi_{\alpha_1}\omega,x)+\alpha_2\varphi_{\alpha_2}(\xi_{\alpha_2}\omega,x)\\
\Psi^{-}(x)&=\beta_1\varphi_{\beta_1}(\xi_{\beta_1}\omega,x)+\beta_2\varphi_{\beta_2}(\xi_{\beta_2}\omega,x),
\end{array} \right.
\end{aligned}
\end{equation}
where  $\varphi_{\alpha_1}$ and $\varphi_{\beta_1}$ correspond to the polarons determined mostly by the bare potentials ($\xi_{\alpha_1}\ge 1$, $0<\xi_{\beta_1} \le 1$ ), $\varphi_{\alpha_2}$ and $\varphi_{\beta_2}$ are tunneling-induced polarons ($0 < \xi_{\alpha_2}\le 1$, $\xi_{\beta_2} \ge 1$). As there are two pairs of polarons, we label this variational ground state by $N=2$. The variational parameters ($\xi_i$, $\alpha_i$ and $\beta_i$) can be determined by minimizing the ground-state energy $E=\bra{G}H\ket{G}$. Details may be found in Appendix \ref{appb}.

 By using this variational state given by Eq. (\ref{twocpolaron}), it is found that the ground-state energy is in high agreement with the exact numerical result for the entire coupling regime, as shown in Fig. \ref{fig2}. The perfect reproduction of the exact numerical result indicates that the ground-state wave function Ansatz given by Eq. (\ref{twocpolaron}) with induced polarons is physically reasonable.

It is also worthwhile to mention that, the relative weight of the induced polaron is certainly related to the tunneling coupling strength $\Omega$, as schematically shown in Fig. \ref{fig1}(d)-(f). The stronger the tunneling strength, the more weight the induced polaron will obatin. This result provides solid support for the tunneling inducing mechanism.

\subsubsection{Bare polaron state in the absence of tunneling}

In order to clarify why the ground state could be constructed in the above way, we start with the exactly solvable case in the absence of tunneling, namely, $\Omega=0$ when the additional potentials are zero. The remaining bare potentials $v^{\pm}$ in Eq. (\ref{barepotential}) accommodate two bare polarons which are bound with two differently shifted modes of harmonics oscillators. Therefore, the profile of these polarons are two Gaussian wavepackets with different frequencies as illustrated in Fig. \ref{fig1}(a). $\Psi^{\pm}$ in  Eq. (\ref{trial}) can be directly given as a deformation of the ground-state wave function $\varphi(\omega,x)$ for an oscillator with the same frequencies as the bare harmonic potentials which differs in opposite spins,
\begin{equation}\label{trial1}
\begin{aligned}
\left\{ \begin{array} {ll}
\Psi^{+}(x)&=\varphi_{\alpha}(\sqrt{(1+2g')/(1-2g')}\omega,x)\\
\Psi^{-}(x)&=\varphi_{\beta}(\sqrt{(1-2g')/(1+2g')}\omega,x)
\end{array} \right.
\end{aligned}
\end{equation}
This wave function is exact in the case of $\Omega = 0$. We call this wave fucntion the bare state. In this exactly solvable case, $\Psi^{+}(x)$ diverges and $\Psi^{-}(x)$ collapses at $g' = \frac{1}{2}$.

If we take this bare state as an analytically approximate result for the $\Omega \neq 0$ case, in the weak coupling regime the energy obtained by this wave function is quite accurate, as shown by the yellow  dash-dotted line in Fig. \ref{fig2}. However, it obviously deviates from the exact numerical result (green circles) for a larger coupling,  and even becomes qualitatively incorrect when $g'\rightarrow \frac{1}{2}$ where an unnormal upward trend is observed. A similar behavior has also been reported in Ref. \cite{1751-8121-49-46-464002} where the tunneling term considered becomes vanishingly small when $g'\rightarrow \frac{1}{2}$. These results indicate that the tunneling process is important in the two-photon QRM, especially near the spectral collapse point.

\subsubsection{Simple variational polaron state (N=1)}

It is reasonable to wonder whether further improvement can be made by simply introducing variationally determined frequency shift parameters $\xi_\alpha, \xi_\beta$  and weights $\alpha, \beta$.
\begin{equation}\label{twopolaron}
\begin{aligned}
\left\{ \begin{array} {ll}
\Psi^{+}(x)&=\alpha\varphi_{\alpha}(\xi_{\alpha}\omega,x),\\
\Psi^{-}(x)&=\beta\varphi_{\beta}(\xi_{\beta}\omega,x).
\end{array} \right.
\end{aligned}
\end{equation}
For convenience, we label this state by $N=1$ indicating that each spin component contains one mode of polaron. This assumption is based on the fact that the effective potentials $v^\pm_\mathrm{eff}$ are asymmetrically deformed from the bare potential $v^{\pm}$ in the case $\Omega \neq 0$, as discussed in Sec. \ref{potential}. For the bare polarons, the first adaptation to the potential deformation is to adjust their frequencies as well as their weights in different spins.

\begin{figure}[tbp]
\includegraphics[width=0.9\columnwidth]{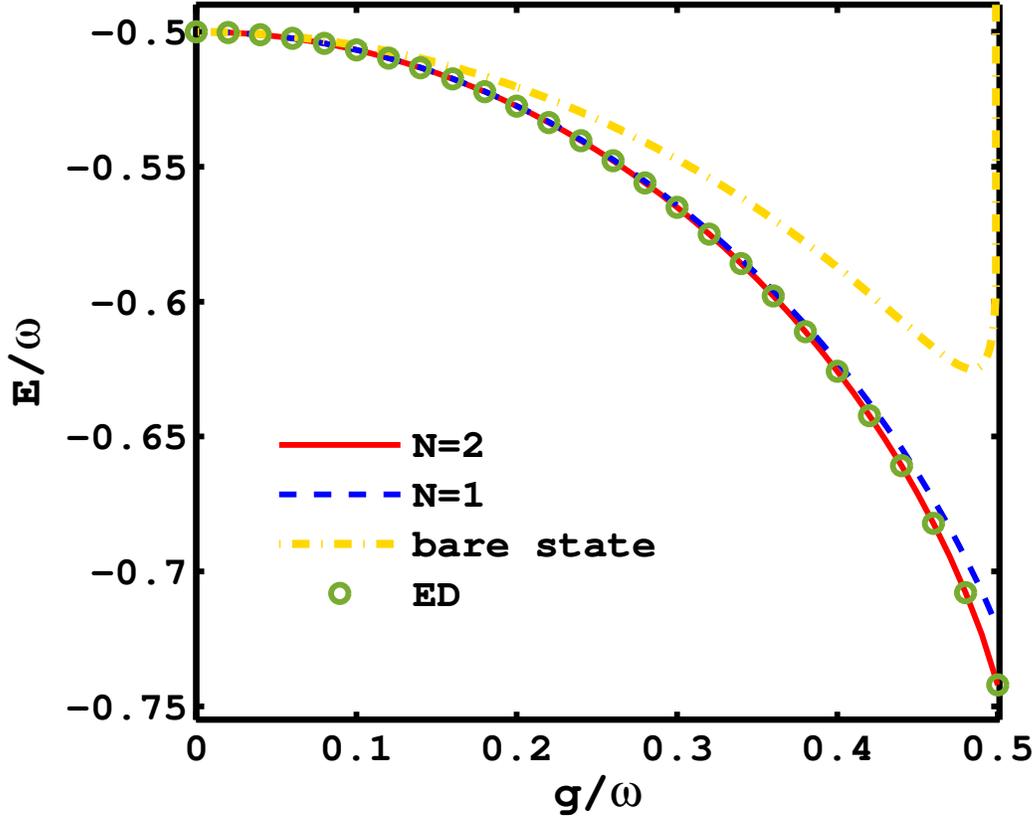}
\caption{(Color online) The ground-state energy as a function of the coupling strength $g$ renormalized by the frequency $\omega$. The red solid line $(N=2)$ represents the result obtained by our frequency-shifted two-pair polaron trial state given by Eq. (\ref{twocpolaron}). The yellow dash-dotted line is the result obtained by the bare state given by Eq. (\ref{trial1}), and the blue dashed line $(N=1)$ denotes the result from the simple trial state given by Eq. (\ref{twopolaron}). In addition, the green circles give the exact diagonalization result which provides a benchmark. Here we take $\omega=1$. For simplicity, we set $\Omega=1$ hereafter.}\label{fig2}
\end{figure}

The ground-state energy obtained is shown in Fig. \ref{fig2} as the blue dashed line. It is noticed that the result has an improvement over the bare state [Eq. (\ref{trial1})]. However, some deviation from the exact numerical calculation is still noticeable in intermediate and strong coupling regimes. This reflects that the simple frequency shift and weight adaptation in Eq. (\ref{twopolaron}) are still not enough to describe the full physics of the tunneling effect. In fact, one will soon find out that the wave function obtained by Eq. (\ref{twopolaron}) is far away from the exact numerical result, which will be discussed in the next section.

\section{Results}\label{results}
\subsection{Wave functions and observables}
In Sec. \ref{method}, we have compared the ground state energy obtained by our polaron picture with the exact numerical calculation. Certainly it is  also worthwhile to check the accuracy of the wave functions as well as other physical properties. In the following, we show explicitly the profile of the ground-state wave function and the behavior of several physical observables including the tunneling amplitude (or spin expectation in $x$ direction) $\langle \sigma_x\rangle$, the mean photon number $\langle a^+a \rangle$ and the correlation $\langle \sigma_z((a^\dag)^2+ a^2)\rangle$.

\begin{figure}[tbp]
\includegraphics[width=0.9\columnwidth]{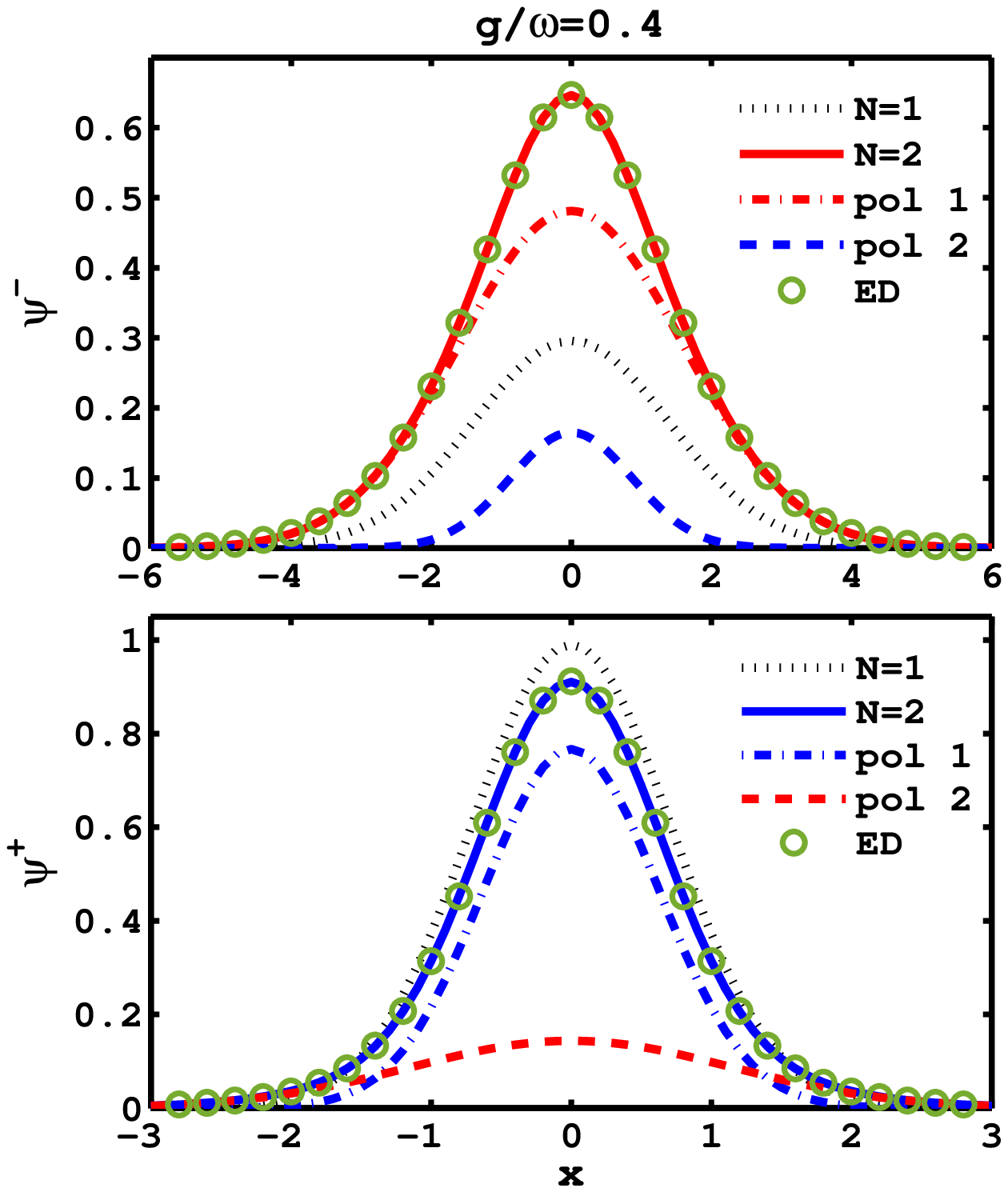}
\caption{(Color online)  The ground-state wave function obtained by the wave function Ansatz with two pairs ($N=2$) of polaron (solid lines), in comparison with the exact numerics (circles) for spin-down (upper panel) and spin-up (lower panel) components. The dash-dotted lines denote original polaron and dashed lines show the polaron induced by the tunneling interaction. The result (dotted lines) for a pair of polarons ($N=1$) is also presented for comparison.}\label{fig3}
\end{figure}

In Fig. \ref{fig3} we plot the ground-state wave function $\Psi^{-} (\Psi^{+})$ and its component(s). As comparison, we also present the exact numerical result (circles). Here, we take $g/\omega = 0.4$, corresponding to the intermediate coupling regime. It is noticed that the wave function $\Psi^{-} (\Psi^{+})$ containing one polaron ($N=1$), denoted by the dotted line in Fig. \ref{fig3}, is far from  the exact numerical result. However, the wave function $\Psi^{-} (\Psi^{+})$ containing two polarons ($N=2$), denoted by the solid lines in Fig. \ref{fig3}, is quite good to reproduce the exact numerical result. In the latter case, $\Psi^{-}$ consists of an original low frequency polaron (pol 1) and a tunneling-induced high frequency polaron (pol 2), as shown in the upper panel of Fig. \ref{fig3}. At the same time, $\Psi^{+}$ comprises an original high frequency polaron (pol 1) and a tunneling-induced low frequency polaron (pol 2), as shown in the lower panel of Fig. \ref{fig3}. The result clearly shows the importance of the induced polarons in capturing the essential physics of the model.

\begin{figure}[tbp]
  \includegraphics[width=0.999\columnwidth]{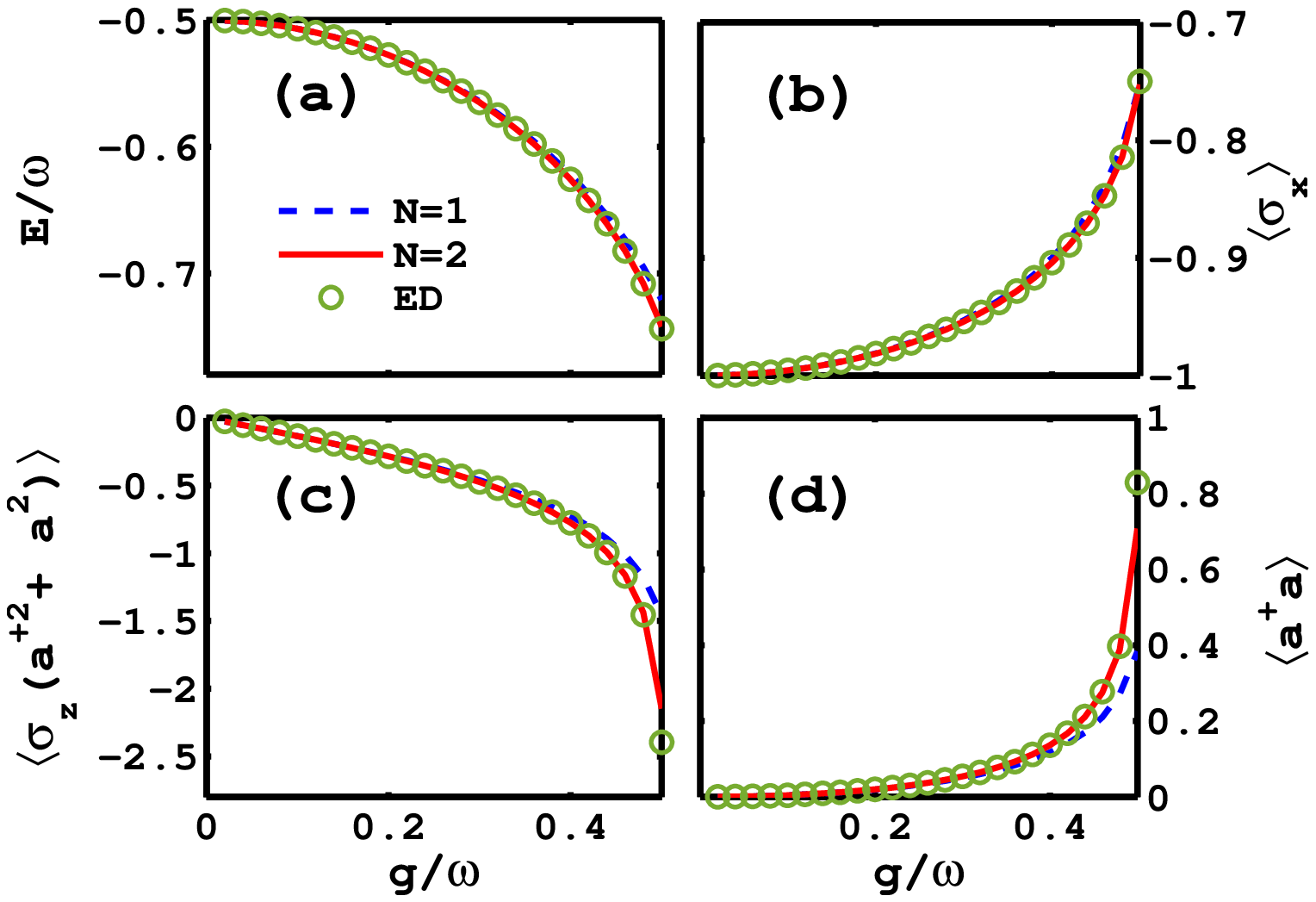}
\caption{(Color online) Physical observables as a function of the coupling strength $g/\omega$. (a) The ground-state energy. (b) The spin polarization $\langle \sigma_x\rangle$. (c) The coupling correlation function $\langle \sigma_z((a^\dag)^2+ a^2)\rangle$. (d) The mean photon number $\langle a^+a \rangle$. The green circles are the exact numerical results taken as benchmarks, and the red solid lines are our results obtained by two pairs of polarons ($N=2$) for the wave function Ansatz Eq. (\ref{twocpolaron}). The results (blue dashed lines) for one pair of ploarons for the wave function Ansatz ($N=1$) is also presented for comparison. Here we take $\omega=1$.}\label{fig4}
\end{figure}

The above comparison in the profile of the wave function illustrates that the accuracy of the variational polaron method may not only lie in the energy but also in the wave function. The latter lays a more essential basis to guarantee the accuracy of the other physical observables. Here we calculate the spin polarization $\langle \sigma_x\rangle$, the correlation function $\langle \sigma_z((a^\dag)^2 +a^2)\rangle$ and the mean photon number $\langle a^\dag a \rangle$, where $\langle\mathcal{O}\rangle= \langle{G} |\mathcal{O}| {G} \rangle$. Fig. \ref{fig4} shows these physical observables as functions of coupling strength $g/\omega$. One notes that our results $N=2$ are in high agreement with the exact numerical diagonalization results, which are obviously better than those obtained by the wave function Ansatz with $N=1$, especially near the spectral collapse point $g'=1/2$.

\subsection{Multi-polaron generalization}\label{generation}

\begin{figure}[tbp]
  \includegraphics[width=0.9\columnwidth]{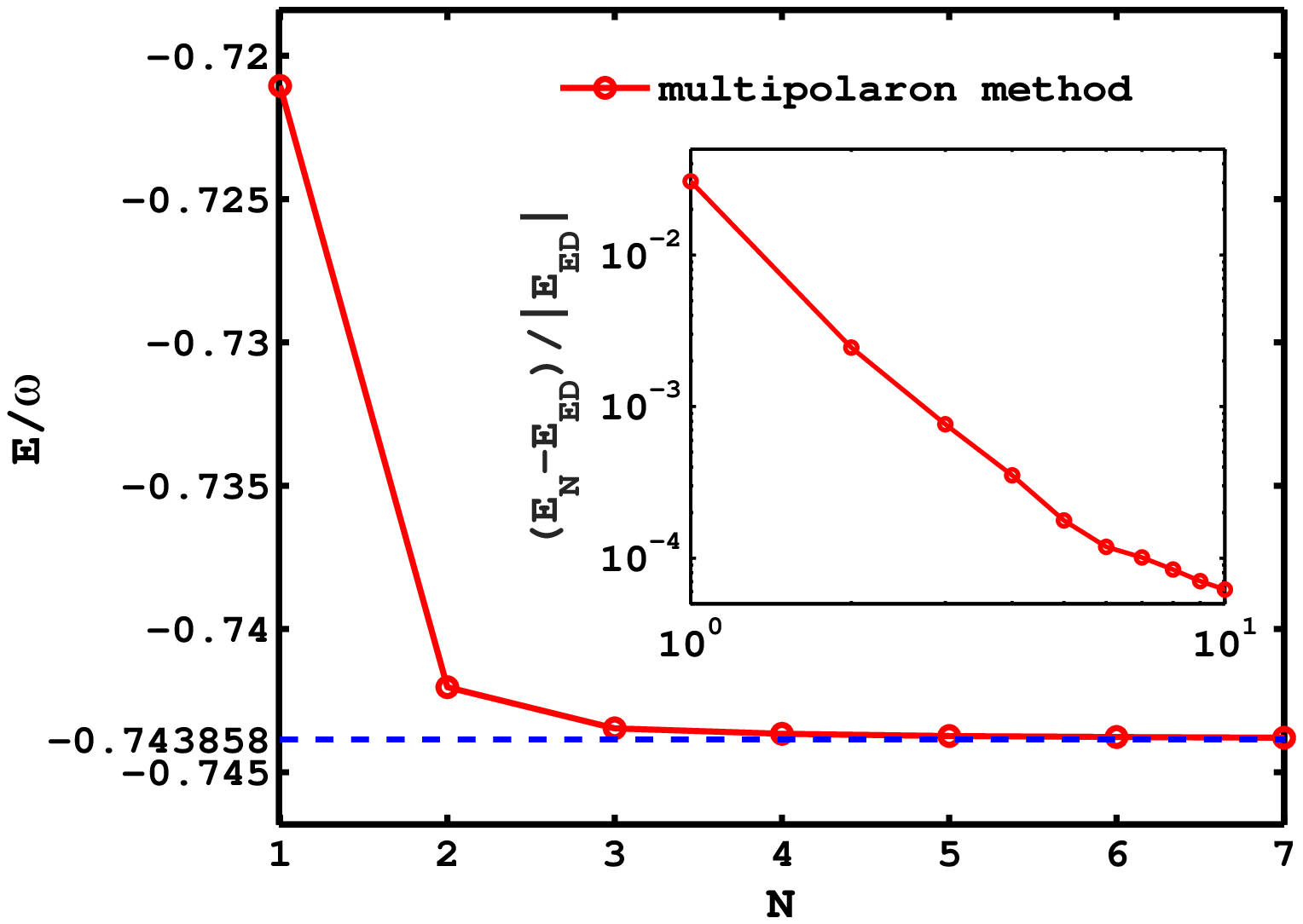}
\caption{(Color online) The ground-state energy as a function of pair number of polarons $N$ at the spectral collapse point ($g/\omega=0.5$). $E_{\rm ED}=-0.743858$ is the energy from exact diagonalization and $E_{\rm N}$ is the energy in multipolaron method. When $N \geq 4$, the error is almost invisible in comparison to the exact numerical result (dashed line). The inset shows the relative error $(E_{\rm N}-E_{\rm ED})/|E_{\rm ED}|$ for the ground-state energy as a function of $N$.}\label{fig5}
\end{figure}

The discussion so far showed that the second polaron induced by tunneling was crucial in the construction of the ground state wave function. However, even though the two-polaron trial state ($N=2$) shows high accuracy, the result can still be further improved by taking higher-order tunneling into account. In this case, the variational wave function can be systematically expanded as

\begin{equation}\label{multipolaron}
\begin{aligned}
\left\{ \begin{array} {ll}
\Psi^{+}(x) &=\sum_{n=1}^N \alpha_n \varphi_{\alpha_n}(\xi_{n}\omega,x),\\
\Psi^{-}(x) &=\sum_{m=1}^N \beta_n \varphi_{\beta_n}(\xi_{n}\omega,x).
\end{array} \right.
\end{aligned}
\end{equation}
Here $N$ denotes the number of the pairs of polarons. It is expected that with $N$ increasing, the results will continue to improve. Notice, the total ground state is defined in Eq. (\ref{trial}), where $\Psi^{\pm}$ is associated with spin state $\uparrow$ and $\downarrow$ to form the polaron. Here, n-th component of $\Psi^{\pm}$, i.e. $\varphi_{\alpha_n/\beta_n}$, when associated with the corresponding spin state $\uparrow / \downarrow$, can be seen as the n-th polaron. Therefore, we call this state a multi-polaron state. For convenience, we sometimes also call $\varphi_{\alpha_n/\beta_n}$ polarons since they are always associated with the corresponding spin state.

In Fig. \ref{fig5} we plot the ground-state energy as a function of $N$  at $g'=0.5$ which is difficult to be solved by other approximate methods as mentioned in the Introduction.  However, using the multipolaron picture, while the two-pair polaron state ($N=2$) discussed in the previous section already reaches a fairly good accuracy, it is noticed that the result converges rapidly with increasing $N$ (almost exponentially, as shown in the inset) and the error for larger $N$ is completely negligible. The result illustrates that the polaron picture indeed works well for the two-photon QRM.

\section{Discussion}\label{Discussion}
In the foregoing, we have seen that the polaron picture provides an effective solution for the ground state properties of the two-photon QRM. In the following we further discuss the polaron picture and point out the difference between the two-photon QRM and one-photon one. In particular, the two-photon QRM shows some novel characteristic features at the spectral collapse point, which have not been explored in the literature.

\subsection{Different polaron pictures in QRM and two-photon QRM}

\begin{figure}[tbp]
  \includegraphics[width=0.9\columnwidth]{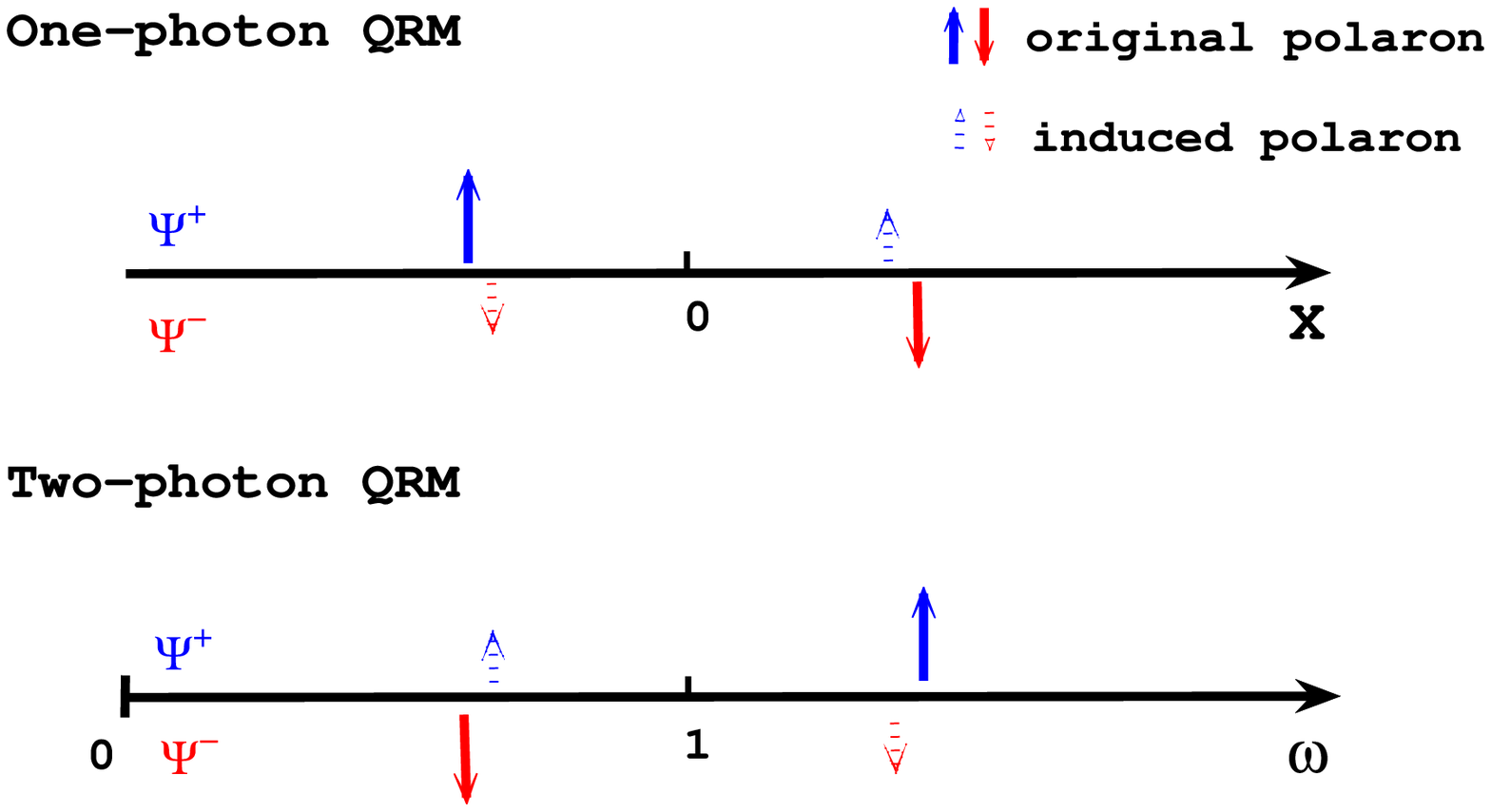}
\caption{(Color online) Schematic comparison of the polaron picture in the one-photon QRM (upper panel) and the two-photon QRM (lower panel). The upper panel shows displaced effective harmonic oscillators, while the lower panel shows frequency shifted effective harmonic oscillators where the original frequency of the bosonic mode is taken to be $1$ as an example.  $\Psi^{+}$ ($\Psi^{-}$) is the spin-up (spin-down) component indicated by the position above (below) the dimension axis. The longer solid (shorter dashed) arrows mark the locations of the original (induced) polarons, with their arrow directions labeling the opposite spins. }\label{figCompareModels}
\end{figure}

The one-photon and two-photon QRM describe linear and nonlinear light-matter interaction, respectively. In the absence of tunneling,  as schematically shown in the upper panel of Fig. \ref{figCompareModels}, the linear interaction leads to position displacement of the bare polarons denoted by solid arrows. For the two-photon QRM, as shown in the lower panel of Fig. \ref{figCompareModels}, the nonlinear interaction leads to a frequency shift of the bare polarons that are also denoted by solid arrows. Based on these features, we can correctly capture the main character of the ground state of these two models by using position displacement or frequency shift parameters, respectively.

When the tunneling is switched on, the new induced polarons will appear, as denoted by dashed arrows.  For both of these two models, the ground state wave-function $\Psi^+(\Psi^-)$ will contain a bare polaron and an induced polaron. This similarity shows that the two models share a common tunneling mechanism.

In order to summarize, the mechanism that tunneling induces additional polarons is the same in these two models, but the linear and nonlinear light-matter interaction manifest the mechanism in different parameter spaces. In the two-photon QRM, a special feature, namely, the spectral collapse will happen since the coupling between light and matter will modify the frequency of the bare polarons, as discussed below.

\subsection{Complete and incomplete spectral collapse}\label{SpectralCollapse}

The most interesting observation in the two-photon QRM is the spectral collapse. Although the spectral collapse has been mentioned previously  \cite{0295-5075-42-1-001,PhysRevA.92.033817,1751-8121-49-46-464002,PhysRevA.95.053854}, it remains elusive, as addressed by Felicetti \textit{et al.} \cite{PhysRevA.92.033817} that it is yet unknown whether the eigenfunctions of $H$ have a plane-wave characteristic or not at $g=\omega/2$. Here, the polaron picture provides a helpful understanding of the spectral collapse.

\begin{figure}[tbp]
  \includegraphics[width=0.99\columnwidth]{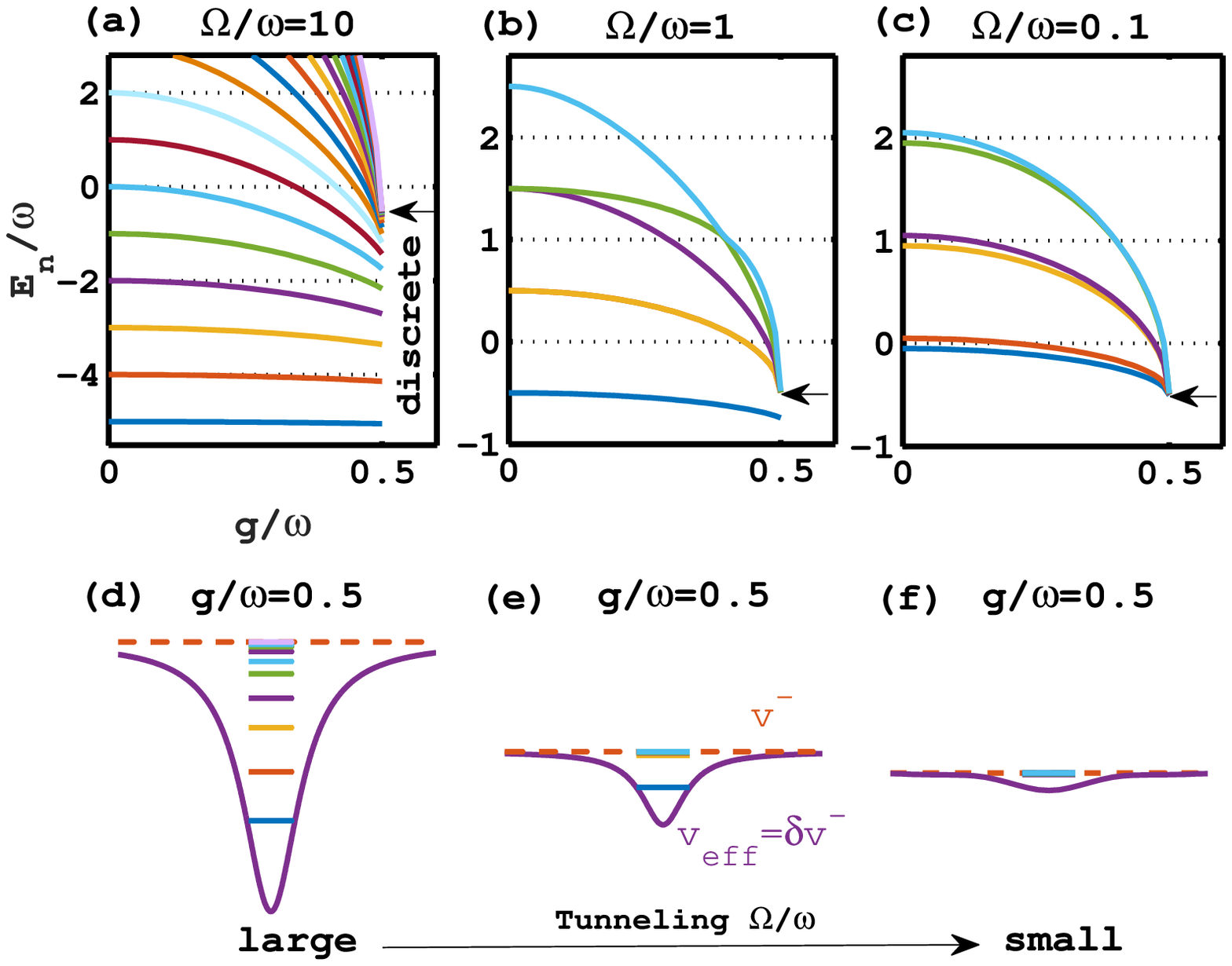}
\caption{(Color online)  (a)-(c) Spectral properties of the two-photon QRM as a function of the coupling strength $g/\omega$ at different tunneling strength $\Omega/\omega$. One can see coexistence of the discrete and collapsed energy levels of the spectrum structure at the collapse point for an finite $\Omega/\omega$. Complete collapse of the spectral happens as $\Omega/\omega \rightarrow 0$. The black solid arrow points to the location of the spectral collapse energy $E_n/\omega=-1/2$. The spectra are calculated by ED with cut off number equals to $1600$, $800$, $800$. (d)-(f) Schematic diagram of the potential curve at $g/\omega=0.5$ where the effective potential would be the tunneling induced potential, $v^-_{eff} =v^-+\delta v^-=\delta v^-$. The depth of the tunneling induced potential well $\delta v^-$ is related to the tunneling strength $\Omega/\omega$. }\label{spectrum}
\end{figure}

In Fig. \ref{spectrum}  (a)-(c), we show the spectrum as a function of coupling strength at different tunneling strengths $\Omega/\omega=10, 1, 0.1$, respectively. When tunneling is strong ($\Omega/\omega=10$), we find that apart from the collapsed energy levels, there also exist some discrete energy levels. Reducing the tunneling strength, the number of the discrete energy levels decreases down to only one as $\Omega/\omega = 1$.  When tunneling is even weaker ($\Omega/\omega=0.1$), it seems all the energy levels collapse to the same point. Moreover, we also notice that the high energy levels collapse to the same energy $E/\omega=-1/2$ at different tunneling strength, as indicated by the black solid arrows in Fig. \ref{spectrum}  (a)-(c).

In the literature, such an incomplete spectral collapse phenomena has not been fully addressed. One notes that Duan \textit{et al.} \cite{1751-8121-49-46-464002} have discussed the gap between ground state energy and the collapsed energy ($E=-1/2$) at $\Omega=\omega=1$ in the framework of the $G$ function, which is the case presented here as Fig. \ref{spectrum}  (b). However, a transparent physical picture is still lacking.

In Fig. \ref{spectrum} (d)-(f), the polaron picture for the spectral collapse point is presented. At this point, the bare potential $v^-$ is flat which is the origin of the collapsed spectrum. However, the tunneling term also contributes an additional potential $\delta v^{-}$, as shown by the solid curve, which is able to localize some discrete bound states. Meanwhile, the highly-excited states will have a plane-wave characteristics because of the finite depth of the effective potential.

The depth of the tunneling induced potential well is strongly dependent on the tunneling strength $\Omega/\omega$ (approximately a linear dependence, see Appendix \ref{appc}). For large tunneling strength $\Omega/\omega$, a deep potential well which locates more discrete energy levels is expected, see Fig. \ref{spectrum}  (d). When $\Omega/\omega \rightarrow 0$, the effective potential tends to be flat and the complete spectral collapse is reached. Meanwhile, in Fig. \ref{spectrum}  (d)-(f), we show an upper limit of the energy for all the energy levels. This energy limit is not zero but the constant $\varepsilon_0=-\omega/2$, as defined below Eq. (\ref{Hamiltonian2}), which sets the final energy value of the collapsing levels.

The physical picture provided here also contributes to the understanding of the results of other methods. For example, it is difficult for the exact diagonalization method to obtain an acceptable converged energy value for the highly-excited states. This is because those states are not well bounded and thus have the plane-wave characteristics. For more details of this argument, see Appendix \ref{appd}.

According to our analysis, even at the spectral collapse point, there is still a tunneling induced potential well, which is able to locate some discrete bound states. Such a polaron picture contributes to the understanding of the incomplete spectral collapse phenomena.

\section{Conclusions} \label{conclusion}

In this paper, we propose a polaron picture for the two-photon QRM. In this picture, without tunneling ($\Omega=0$), it is found that the nonlinear light-matter couping splits the single bosonic mode into two different modes corresponding to the spin up and down states, respectively. These two modes are depicted as two bare potentials $v^{\pm}$ which then form two \textit{bare} polarons with different frequency. As $\Omega\neq0$, the bare potentials $v^{\pm}$ are deformed by the tunneling induced additional potentials $\delta v^{\pm}$. Intuitively, the two bare polarons exchange their components through the tunneling process, leading to \textit{induced} polarons. Thus, a variational ground-state wave function Ansatz with two pairs of polarons has been proposed, which is found in good agreement with the numerical exact result. The physical observables calculated from the polaron picture are also consistent with the numerical results in the entire coupling regime. Furthermore, the method is generalized to $N$ pairs of polarons showing a fast convergence of accuracy with increasing N.

The polaron picture provides a helpful understanding of the spectral collapse.  We found that even at the spectral collapse point, there is still a tunneling induced potential well, which is able to locate some discrete bound states. Thus, the spectral collapse is not complete for finite tunneling. Only when  $\Omega/\omega$ approaches zero, the spectral collapse becomes complete due to a quite shallow potential well. Either in complete or incomplete spectral collapse, the energy limit of the collapse part is given by the constant energy $\varepsilon_0=-\omega/2$ figuring in the polaron method.

Further applications and investigations on the related two-photon models are also possible, for example, the biased two-photon QRM, the two-photon Dicke model \cite{PhysRevA.95.053854}, the two-photon spin-boson model where a quantum spin nonlinearly couples to a bosonic bath. These and similar models \cite{Casanova2018, 1751-8121-50-29-294004} may be studied using the polaron method in the future.

\section*{Acknowledgments}
 We acknowledge useful discussion with Simone Felicetti, Hans-Peter Eckle, Elinor Irish, Yu-Yu Zhang, Li-Wei Duan, Gao-Yang Li, Fu-Zhou Chen and Yuan-Sheng Wang. In particular, we thank Hans-Peter Eckle for his helpful on the presentation of whole text. This work was supported by NSFC (Grant No. 11674139, 11834005, No. 11604009), PCSIRT (Grant No. IRT-16R35) and the Fundamental Research Funds for the Central Universities. Z.-J.Y. acknowledges the financial support of the Future and Emerging Technologies (FET) programme within the Seventh Framework Programme for Research of the European Commission, under FET-Open Grant No. 618083 (CNTQC).

\appendix
\section{Numerical ground state wave functions}\label{appa}

\begin{figure}[tbp]
  \includegraphics[width=0.99\columnwidth]{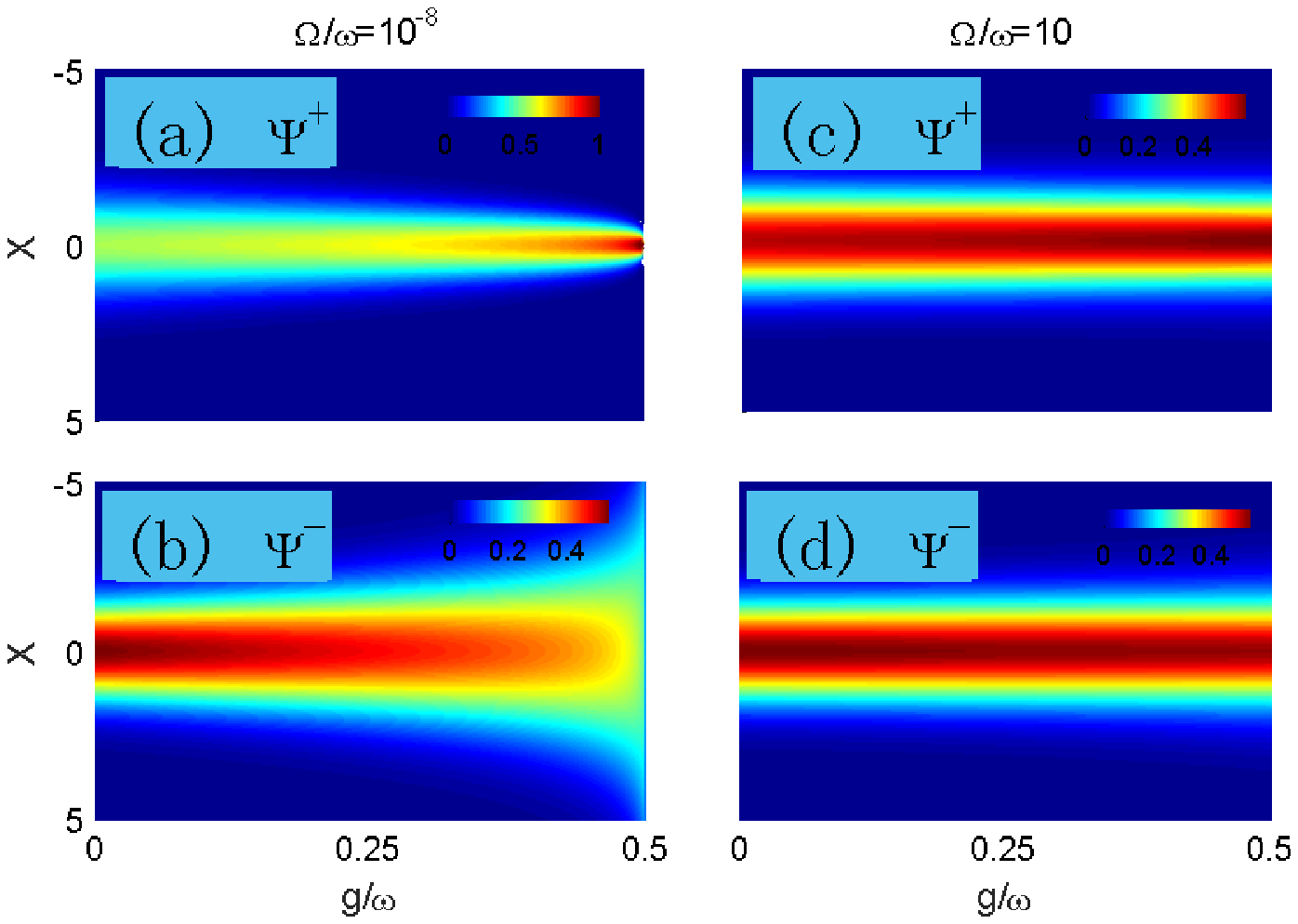}
\caption{(Color online) An overview of the ground state wave functions $\Psi^{\pm}(x,g/\omega)$. Here, $x$ is the position and $g/\omega$ is the coupling strength. Left panel: Tunneling strength $\Omega/\omega \rightarrow 0$. The ground state property is determined by the bare potential $v^{\pm}$. Thus, $\Psi^+$ will diverge and $\Psi^-$ will spreads out as $g/\omega$ increases. Right panel: Tunneling strength $\Omega/\omega =10$. Large tunneling make $\Psi^{\pm}$ neutralize each other.}\label{wavecompare}
\end{figure}

Fig. \ref{wavecompare} provides the ED wave packets $\Psi^{\pm}(x)$ as a function of the coupling strength $g/\omega$ (with a cut off number 800). In the case $\Omega/\omega \rightarrow 0$, wave function $\Psi^+(x)$ becomes sharper and sharper as the coupling strength increases, see Fig. \ref{wavecompare} (a). Meanwhile, $\Psi^-(x)$ continuously spreads out and collapsed finally, see Fig. \ref{wavecompare} (b). However, in the case of  $\Omega/\omega =10$, the large tunneling changes the system properties completely. The trend to diverge and collapse of $\Psi^{\pm}(x)$ are neutralized by the tunneling induced component of the wave function. Therefore, $\Psi^{+}(x)$ and $\Psi^{-}(x)$ become very similar,  see the right panel of Fig. \ref{wavecompare}.

\section{The ground-state energy of multi-polaron state}\label{appb}

In this appendix, we present the main steps to get the ground-state energy.
\subsection{Trial wave function}
The wave function of the model can be generally written as
\begin{equation}\label{eq:eps}
\ket G=\frac{1}{\sqrt{2}}\big(\Psi^{+}(x)\ket\uparrow-\Psi^{-}(x)\ket\downarrow \big).
\end{equation}
where $+$ and $\uparrow$ label the spin-up component while $-$ and $\downarrow$ label the spin-down component.
In the multi-polaron states, each spin component can be expanded as
\begin{equation}\label{eq:eps}
\begin{aligned}
\left\{ \begin{array} {ll}
\Psi^{+}(x)&=\sum_{n=1}^N \alpha_n \varphi_{\alpha_n}(x),\\
\Psi^{-}(x)&=\sum_{n=1}^N \beta_n \varphi_{\beta_n}(x),\\
\end{array} \right.
\end{aligned}
\end{equation}
For one pair or two pairs of polarons, one simply choose $N=1$ or $N=2$. Here, $\alpha_i$ and $\beta_i$ are the corresponding polaron weights and $\varphi$ is a variational version of the ground state of the quantum harmonic oscillator $\phi_0$ with frequency shift factor $\xi_i$ inside
\begin{equation}\label{eq:eps}
\begin{aligned}
\left\{ \begin{array} {ll}
\varphi_{\alpha_n}(x)&=\phi_0(\xi_{\alpha_n}\omega,x),\\
\varphi_{\beta_n}(x)&=\phi_0(\xi_{\beta_n}\omega,x),\\
\end{array} \right.
\end{aligned}
\end{equation}
and
\begin{equation}\label{eq:eps}
\phi_0(\omega,x)=(x_0\sqrt{\pi})^{-\frac{1}{2}}e^{-\frac{x^2}{2{x_0}^2}},
\end{equation}
where $x_0^2=\hbar/{m\omega} $ and we take $\hbar=1,m=1$ for simplicity. In terms of the quantum harmonic oscillator representation in the dimensionless formalism, we have for $\varphi$ explicitly
\begin{equation}\label{eq:eps}
\left\{ \begin{array} {ll}
\begin{aligned}
\phi_0(x) &=(\frac{\xi_{\alpha_n}}{\pi})^{\frac{1}{4}}e^{-\frac{x^2\xi_{\alpha_n}}{2}},\\
\phi_0(x) &=(\frac{\xi_{\beta_n}}{\pi})^{\frac{1}{4}}e^{-\frac{x^2\xi_{\beta_n}}{2}}.\\
\end{aligned}
\end{array} \right.
\end{equation}

\subsection{Derivation of the ground-state Energy}

With the wave function in multipolaron form, it is easy to calculate the energy
\begin{equation}\label{eg}
\begin{aligned}
E & =\bra{G}H\ket{G}\\
&=\frac{1}{2} \big( \bra{\Psi^{+}} h^{+} \ket{\Psi^{+}} +\bra{\Psi^{-}} h^{-} \ket{\Psi^{-}} \big) \\
&-\frac{\Omega}{2} \big( \frac{1}{2} \langle{\Psi^{+}} | {\Psi^{-}} \rangle+ \frac{1}{2} \langle{\Psi^{-}} |{\Psi^{+}} \rangle \big)+\varepsilon_0.\\
\end{aligned}
\end{equation}
Each term in Eq. (\ref{eg}) can be obtained respectively as follows, with the single-particle energy in the absence of tunneling,
\begin{equation}\label{eq:eps}
\begin{aligned}
h_{++}^{+}&=\bra{\Psi^{+}}h^{+}\ket{\Psi^{+}}\\
&= \sum_{n,m} \alpha_n \alpha_m \bra{\varphi_{\alpha_n}}h^{+}\ket{\varphi_{\alpha_m}}\\
&= \omega/2\sum_{n,m} \alpha_n \alpha_m \bra{\varphi_{\alpha_n}}(1-2g^{'})p^2+(1+2g^{'})x^2\ket{\varphi_{\alpha_m}}\\
\end{aligned}
\end{equation}
and
\begin{equation}\label{eq:eps}
\begin{aligned}
h_{--}^{-}&=\bra{\Psi^{-}}h^{-}\ket{\Psi^{-}}\\
&= \sum_{n,m} \beta_n \beta_m\bra{\varphi_{\beta_n}}h^{-}\ket{\varphi_{\beta_m}}\\
&= \omega/2\sum_{n,m} \beta_n \beta_m\bra{\varphi_{\beta_n}}(1+2g^{'})p^2+(1-2g^{'})x^2\ket{\varphi_{\beta_m}},\\
\end{aligned}
\end{equation}
as well as the wave-packet overlap for the tunneling term
\begin{equation}\label{eq:eps}
\begin{aligned}
h_{+-}&=\langle{\Psi^{+}}|{\Psi^{-}}\rangle= \sum_{n,m} \alpha_n \beta_m\langle{\varphi_{\alpha_n}}|{\varphi_{\beta_m}}\rangle \\
&= \omega/2\sum_{n,m} \alpha_n \beta_m\langle{\varphi_{\alpha_n}}|{\varphi_{\beta_m}}\rangle \\
\end{aligned}
\end{equation}
and $h_{-+}$ is equal to $h_{+-}$.
The overlap is explicitly available as
\begin{equation}\label{eq:overlap}
\begin{aligned}
S_{\alpha_n\beta_m}&=\langle{\varphi_{\alpha_n}}|{\varphi_{\beta_m}}\rangle=\sqrt{2}\big[ \frac{\xi_{\alpha_n}\xi_{\beta_m}}{(\xi_{\alpha_n}+\xi_{\beta_m})^2}\big]^{\frac{1}{4}}
\end{aligned}
\end{equation}
and similarly we get the potential term
\begin{equation}\label{eq:eps}
\begin{aligned}
{\langle{\hat{x}}^2\rangle}_{\alpha_n\beta_m}&=\bra{\varphi_{\alpha_n}} {\hat{x}}^2 \ket{\varphi_{\beta_m}}=S_{\alpha_n\beta_m} (\frac{1}{\xi_{\alpha_n}+\xi_{\beta_m}}) \\
\end{aligned}
\end{equation}
and the kinetic contribution
\begin{equation}\label{eq:eps}
\begin{aligned}
\bra{\varphi_{\alpha_n}}(-\frac{\partial^2{}} {\partial{x^2}}) \ket{\varphi_{\beta_m}}&=-\bra{\varphi_{\alpha_n}} \xi_{\beta_m}^2 x^2 -\xi_{\beta_m}\ket{\varphi_{\beta_m}}.\\
\end{aligned}
\end{equation}
The ground-state energy is determined by the minimization of the energy with respect to the variational parameters of $\xi_i$, $\alpha_i$ and $\beta_i$, subject to normalization condition as follows,
\begin{equation}\label{eq:eps}
\begin{aligned}
\langle{G}|{G}\rangle&=\frac{1}{2} (\langle{\Psi^{+}}|{\Psi^{+}}\rangle+\langle{\Psi^{-}}|{\Psi^{-}}\rangle)\\
&=\frac{1}{2}( \sum_{n,m} \alpha_n \alpha_m\langle{\varphi_{\alpha_n}}|{\varphi_{\alpha_m}}\rangle +\sum_{n,m} \beta_n \beta_m\langle{\varphi_{\beta_n}}|{\varphi_{\beta_m}}\rangle)\\
&= 1,
\end{aligned}
\end{equation}
where the overlaps are ready in \eqref{eq:overlap}.

\section{Properties of the spectum at the collapse point}\label{appc}

\begin{figure}[tbp]
  \includegraphics[width=0.99\columnwidth]{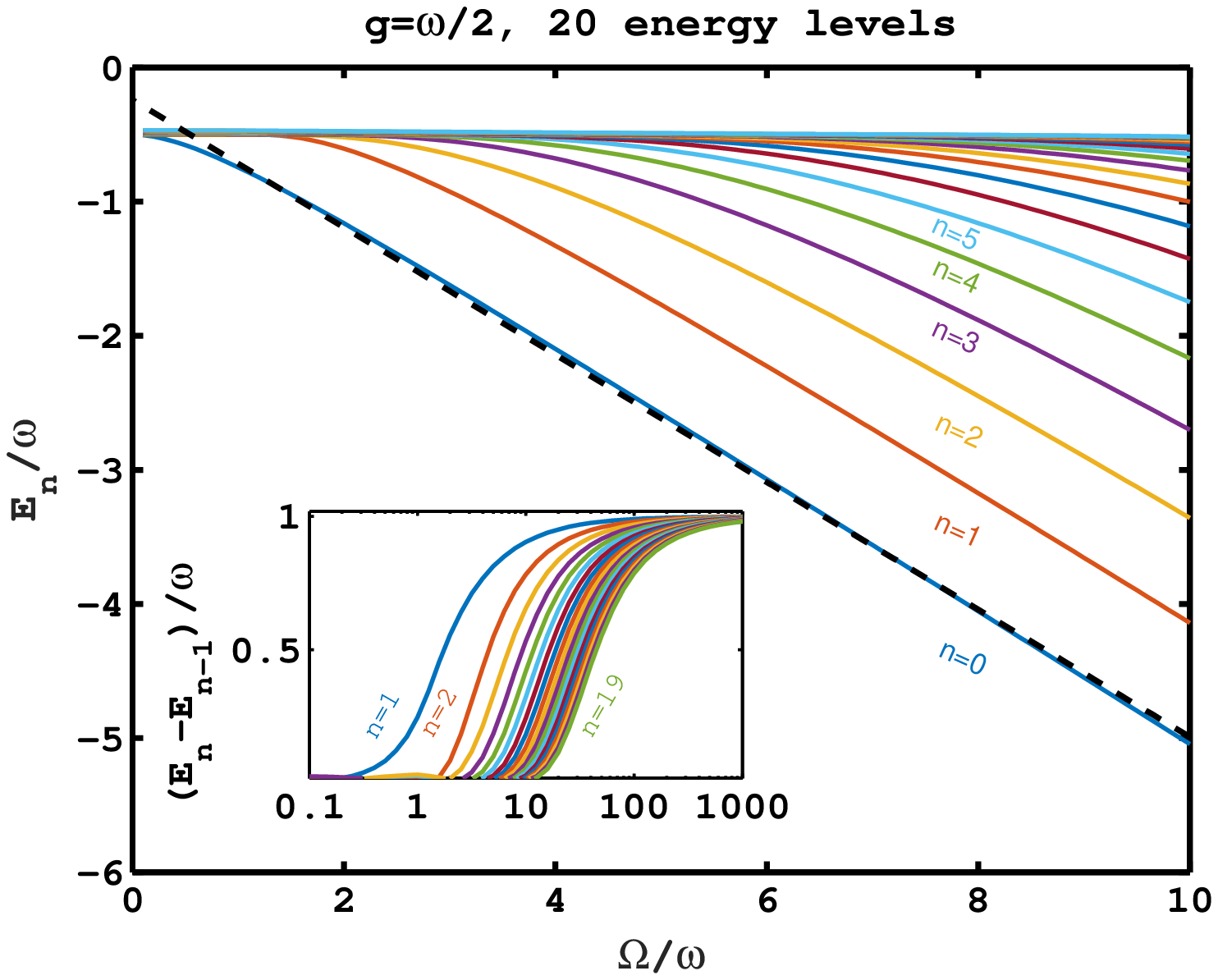}
\caption{(Color online) Spectral properties of the two-photon QRM as a function of the tunneling strength $\Omega/\omega$. The inset shows the energy level intervals changes as a function of the tunneling strength $\Omega/\omega$. }\label{spectrum3}
\end{figure}

Fig. \ref{spectrum3} provides the ED spectrum at the collapse point as a function of the tunneling strength $\Omega/\omega$. It is noticed that more and more energy levels depart from the collapsed spectrum energy $E_n/\omega=-1/2$ as $\Omega/\omega$ increases. This is because the depth of the tunneling induced potential well strongly depends on the tunneling strength. We can see that the ground state energy has an approximate linear relation with $\Omega/\omega$, as indicated by the dashed line. Besides, at large tunneling, the energy level intervals trends to $1$ as shown in the inset. This is a characteristic of the harmonic oscillator potential well.

\section{An understanding from the polaron picture for the spurious ED convergence above the spectral collapse}\label{appd}

\begin{figure}[tbp]
  \includegraphics[width=0.99\columnwidth]{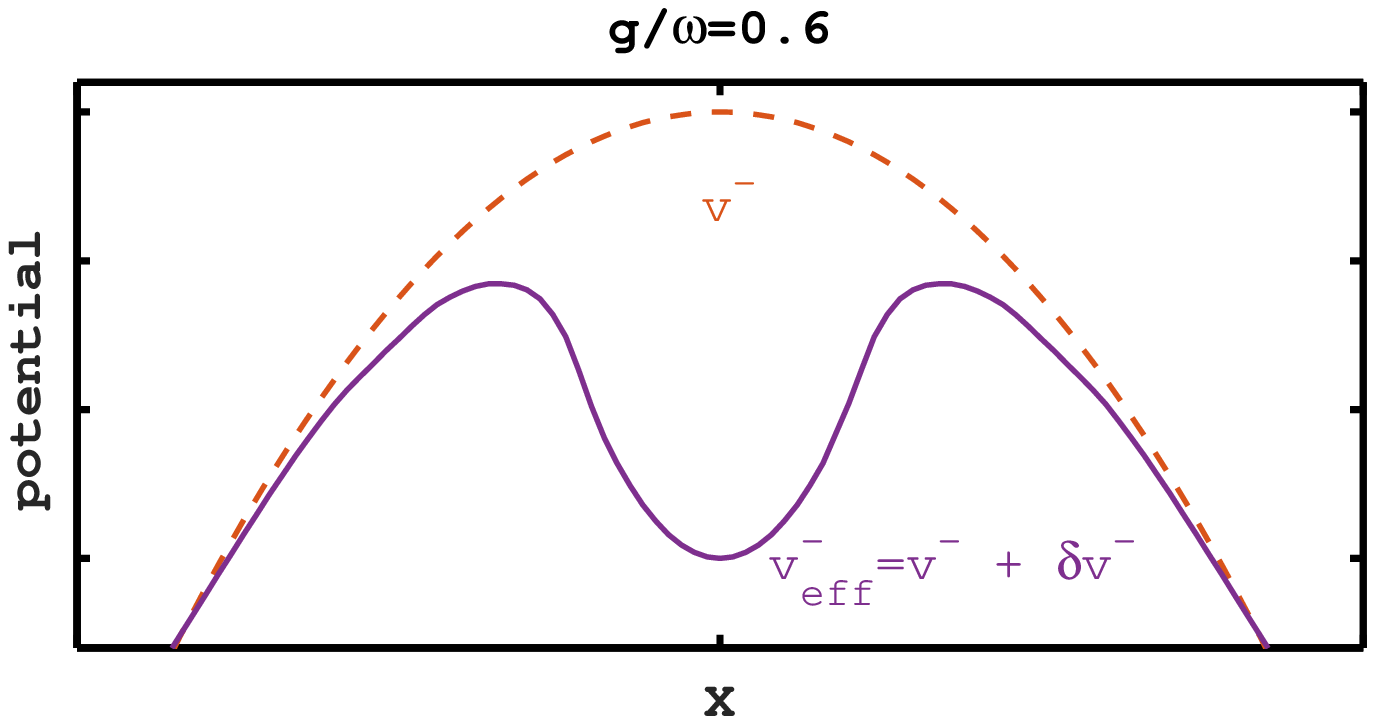}
\caption{(Color online) Schematic diagram  of the potential curve above the spectral collpase point. Tunneling process also contributes an additional potential well which may lead to a spurious convergence in ED. }\label{potentialfig}
\end{figure}

\begin{table}[!hbp]
\caption{Energy dependence on the cutoff for the ground state $E_0/\omega$ and the first excited state $E_1/\omega$ at large tunneling case ($\Omega/\omega=1000$) above the spectrum collapse point ($g/\omega=0.6$). }
\begin{tabular}{c   c   c}
\hline
\hline
cutoff & $E_0/\omega$ & $E_1/\omega$ \\
\hline
400 & -50.0000719339646& -49.9002159576017 \\

800 & -50.0000719339646&  -49.9002159576017 \\

1200 & -50.0000719339645& -49.9002159576016 \\

1600 & -50.0000719339646&  -49.9002159576017 \\

2000 & -50.0000719339646&  -49.9002159576017 \\

2400 & -51.0880505329409&   -51.0880483452417   \\

2800 & -58.3843972619404&  -58.3843956108931  \\

3200 & -65.8437067512263& -65.8437054653073 \\

3600 & -73.4130541896708& -73.4130531617850  \\

4000 & -81.0602940889838& -81.0602932495258\\
\hline
\hline
\end{tabular}\label{table1}
\end{table}

For $g>\omega/2$, the potential will become a barrier if only the bare potential part is considered. Thus, in this regime, the system would not have bound states.  However, it is usually encountered an artifact for the ED that the energy appears to be convergent at first, whereas with a larger cutoff the ED method breaks down completely, see Table. \ref{table1}. Similar behavior is also addressed in the literature \cite{0295-5075-42-1-001}.

A barrier potential can not explain the first false convergence.  In Fig. \ref{potentialfig}, as discussed in the polaron picture, we further take the induced potential into account. We see that the total potential (solid line) has a local well in the top of the barrier. One immediately recognizes that, for the ED method, the induced local potential well is felt first in a finite cutoff. Thus the numerical result shows a spurious convergence. When the cutoff is large enough to detect the global barrier which has no lower bound, the ED will not become convergent as the cutoff further increases. Thus, from this picture it is understand the confusion about the convergence behavior in the ED above the spectral collapse, and we see that in such a situation the numerical convergence in ED observed in small cutoff matrices is spurious and not reliable.

\bibliography{paper}

\end{document}